\documentclass{aa}

\usepackage{graphicx}
\usepackage{txfonts}
\usepackage{epsfig}

\begin{document}
\title{
Discovery of New Milky Way Star Cluster Candidates in the 2MASS 
Point Source Catalog II.\\ 
Physical Properties of the Star Cluster CC\,01}
\subtitle{}

\author{J.~Borissova\inst{1}
\and
P.~Pessev\inst{2}
\and
V.D.~Ivanov\inst{3}
\and 
I.~Saviane\inst{4}
\and
R.~Kurtev\inst{2}
\and 
G.R.~Ivanov\inst{2}
}

\offprints{J.~Borissova}

    \institute{
	Pontificia Universidad Cat\'{o}lica de Chile, Facultad de F\'{\i}sica, 
	Departamento de Astronom\'{\i}a y Astrof\'{\i}sica,
	Av. Vicu\~{n}a Mackenna 4860, 782-0436 Macul, Santiago, Chile\\
	\email{jborisso@astro.puc.cl}
      \and
        Department of Astronomy, Sofia University, Bulgaria, and
        Isaac Newton Institute of Chile, Bulgarian Branch,
        5~James Bourchier, 1164 Sofia, Bulgaria\\
        \email{pessev, kurtev, givanov@phys.uni-sofia.bg}
      \and
        European Southern Observatory, Karl-Schwarzschild-Str. 2,
	D-85748 Garching bei Mnchen, Germany\\
        \email{vivanov@eso.org}
      \and
        European Southern Observatory, Ave. Alonso de
        Cordova 3107, Casilla 19, Santiago 19001, Chile\\
        \email{isaviane@eso.org}
      }

\date{Received .. ... 2003; accepted .. ... 2003}

\authorrunning{Borissova et al.}
\titlerunning{New Milky Way Star Clusters II}

\abstract{
Three new obscured Milky Way clusters were detected as surface 
density peaks in the 2MASS point source catalog during our on-going
search for hidden globular clusters and massive Arches-like star 
clusters. One more cluster was discovered serendipitously during a 
visual inspection of the candidates. 

The first deep $J$, $H$, and $K_s$ imaging of the cluster 
[IBP\,2002]\,CC\,01 is presented. We estimated a cluster age of 
$\sim$1-3 Myr, distance modulus of (m-M)$_0$=12.56$\pm$0.08 mag 
(D=3.5 Kpc), and extinction of A$_V$$\sim$7.7 mag. We also derived 
the initial mass function slope for the cluster: $\Gamma$=$-$2.23$\pm$0.16
($\Gamma$$\rm_{Salpeter}$=$-$2.35). The integration over the initial 
mass function yielded a total cluster mass 
$M_{total}$$\leq$1800$\pm$200$M_\odot$. CC\,01 appears to be a regular, 
not very massive star cluster, whose formation has probably been 
induced by the shock front from the nearby H{\sc ii} region 
Sh\,2-228.

\keywords{(Galaxy:) open clusters and associations: general - Infrared:
general}

}

\maketitle

\section{Introduction}

All-sky infrared surveys offer the opportunity to study for the first 
time the obscured cluster population in the disk of the Milky Way that 
is hidden by severe dust extinction ($\rm A_V$=10-20 mag). A major 
drive for such studies are the discovery of new globular clusters (i.e. 
Hurt et al. \cite{hur00}) and massive ($\lesssim 7\times10^4~M_\odot$)
Arches-like
young clusters (Nagata et al. \cite{nag93}). While the former offer 
the chance to reveal the full extent of the old stellar population 
in the Milky Way disk -- the so-called disk globulars, the latter
help to probe the most extreme star formation environments, together 
with more traditional problems such as the spiral structure of the Galaxy
and the abundance gradients.

A number of searches have already been undertaken and the most recent 
catalogs contain about 500 newly discovered objects (Bica et al. 
\cite{bic03}). The techniques vary from simple visual inspection of 
regions with known mid-infrared or radio emission (Dutra et al. 
\cite{dut02}) to fully automated density variation searches (Ivanov 
et al. \cite{iva02}, hereafter Paper {\sc I}; Reyl\'{e} \& Robin 
\cite{rey02}). The former ones take advantage of the radio- and 
mid-infrared emission of young clusters, and the later are sensitive 
to all clusters regardless of their age and population. Our on-going 
project is aimed at discovery of previously unknown Galactic globular 
clusters or/and massive Arches-like young clusters. 

Here we present a set of new clusters. This paper concludes our search 
with 5$\times$5 arcmin bin size which determines 
the scale of the objects. Next, we will concentrate on smaller structures. 

We also report a mass estimate of [IBP\,2002]\,CC\,01 (hereafter CC\,01) 
that was selected from the list in Paper {\sc I} as a massive cluster 
candidate. This study is a demonstration of the follow up we will carry 
out on the most promising clusters. Fortuitously, the two brightest 
stars in CC\,01 have optical spectroscopy available in the literature 
that allows us to determine the distance and the cluster mass. However, 
we intend to use near infrared spectroscopy for the objects with higher 
obscuration.

\section{New Cluster Candidates}

\subsection{Search Technique and Strategy}

Our search technique is based on local fluctuations in the stellar 
surface density. The first step is to build a 2-dimensional histogram 
in terms of number of stars per unit area on the sky. Next, the value 
in each bin is compared with the background, calculated from the 
surrounding bins. The bin size determines the characteristic size of the 
selected structures.  
The main advantage of the surface density method is that it is sensitive 
to all populations of clusters. Our strategy was to search first for 
larger cluster candidates. Therefore, we started with 5$\times$5 arcmin 
bins. For more detailed description see Paper {\sc I}.

This technique has one caveat - the density peaks produced by objects
close to the bin borders can fall into two or more bins. Such peaks 
could be diluted beyond detection. To alleviate this problem we repeated 
the search shifting the centers of the bins by 2.5 arcmin along both 
axis. The present work is focused on the regions with Galactic latitude 
$b\leq$10 degr, because our previous experience showed that only 
clusters hidden behind A$_V\sim10-20$ mag have evaded previous detection. 
We plan to decrease the size of the bins in our future searches, 
optimizing it to detect smaller objects.

\section{New Clusters\label{IndivObj}}

Parameters of the new cluster candidates are given in 
Table~\ref{TblCandidates}. Three of them were discovered by our 
searching software, and one - during visual inspection of the 
candidates. The numbering continues that in Paper {\sc I} and the
"CC" means "cluster candidate". Here we 
refrained from numerical estimates of the total cluster apparent 
luminosity because our experience with CC\,01 showed that the cluster 
boundaries depend on the depth of the images, and the magnitudes 
determined from the 2MASS images are underestimated. The 
diameters are approximate eye-ball estimates from the $K_s$ band images. 
True-color reproductions of the 2MASS images of the new clusters are 
shown in Fig.~\ref{figRGB}.

\begin{table}[t]
\begin{center}
\caption{Parameters of the cluster candidates. The first three
objects were identified by the automatic algorithm, and the last 
one was found serendipitously after a visual inspection. See
Sec.~\ref{IndivObj} for comments on individual objects.}
\label{TblCandidates}
\begin{tabular}{l@{ }c@{ }r@{}c@{}}
\hline
\multicolumn{1}{c}{ID} &
\multicolumn{1}{c}{R.A.  Dec.} &
\multicolumn{1}{c}{{\it l   b}} &
\multicolumn{1}{c}{D} \\
\multicolumn{1}{c}{CC} &
\multicolumn{1}{c}{(J2000.0)} &
\multicolumn{1}{c}{} &
\multicolumn{1}{c}{$\arcmin$} \\
\hline
 11 & 06:33:27 $+$12:03.5 & 200.17 $+$1.55 & 0.6 \\ 
 12 & 08:16:31 $-$35:40.0 & 253.72 $-$0.21 & 0.5 \\ 
 13 & 20:31:34 $+$45:05.8 &  83.09 $+$3.28 & 1.0 \\ 
\hline
 14 & 05:28:59 $+$34:23.2 & 173.50 $-$0.06 & 1.0 \\ 
\hline
\end{tabular}
\end{center}
\end{table}
\begin{figure}
  \caption{Near infrared 3-color composite image of the new 
  cluster candidates. The $J$, $H$, and $K_s$ bands are 
  mapped onto blue, green and red, respectively. From left to 
  right, and from top to bottom:  IR cluster candidate No69 
  (Bica et al \cite{bic03b}), CC\,11, CC\,12, CC\,13, 
  CC\,14. The images is 4 arcmin on the side, except 
  for CC\,14, which is 4x9 arcmin. North is up, and East is to 
  the left. Some artifacts -- mainly reflections from bright 
  nearby stars -- are seen as monochromatic dots, particularly
  in the case of CC\,13. The 
  IR cluster candidate No69 shown for comparison 
  was also found with our algorithm.
}
  \label{figRGB}
\end{figure}

Comments to individual objects:

{\bf CC\,11}: The objects appears as a compact group of reddened 
stars ($J-K_s\leq$1.5 mag) on the color-magnitude diagram. Some
stars are so obscured that they are visible on the $K_s$ image 
only. Extended emission is present indicating that this might be 
a dust-rich young cluster.

{\bf CC\,12}: A compact group, dominated by few bright stars. It 
is located $\sim$0.5 arcmin away from the infrared source 
IRAS\,08146-3529A. 

{\bf CC\,13}: This object is in the outskirts of the Cygnus 
Complex which together with the presence of a molecular cloud 
([DBY94] 083.1+03.3; Dobashi et al. \cite{dob94}) and two 
radio-sources (F3R\,2728, MITG\,J2031+4505) makes it a likely 
young cluster. Note that the radio sources can well be 
misidentified as a single one. The spatial distribution of stellar 
colors indicates that the cluster is embedded into a large 
($>$10 arcmin) long dust lane.

{\bf CC\,14}: The cluster was serendipitously discovered in the
field of CC\,11 as a round group of red stars. The over-density is 
bellow our detection limit but the stellar colors are well 
distinguished on the color-magnitude diagram. The object has two 
tidal-tail-like structures pointing in opposite directions that 
can be traced out as far as $\sim$3 arcmin to the East and $\sim$5 
arcmin to the West.
Two nearby mid-IR sources are located in the vicinity of the 
cluster: IRAS\,05257+3420 at $\sim$1.5 arcmin, and IRAS\,0553+3421 
at $\sim$2.5 arcmin. However there is no evidence of physical 
association. 
Deeper photometry is necessary to reveal the true 
nature of the object.

\section{Physical Parameters of CC\,01}

The cluster candidate CC\,01 ($\alpha_{2000}$=05:13:26 
$\delta_{2000}$=+37:27.0) was selected as a part of our on-going 
search (Paper {\sc I}). It was chosen for this follow up study 
because it appeared richer and more massive than the rest of our 
candidates.

Some cluster members were detected in a study of embedded star 
clusters by Carpenter et al. (\cite{car93}). They are associated 
with a H{\sc ii} region Min\,2-58 (= Sh\,2-228; 
Minkowski \cite{min48}, Sharples \cite{sha59}). Optical photometry 
of the brightest four stars in the H{\sc ii} region was reported 
by Lahulla (\cite{lah85}) who determined the spectral types of two 
stars from their colors. Later, Hunter \& Massey (\cite{hun90}) and
Hunter (\cite{hun92}) obtained optical spectra of two bright stars,
and estimated that the H{\sc ii} region is at 2.4 Kpc from the Sun, 
based on these spectral types, and assuming the stars are on the 
main sequence. They also determined the reddening toward the cluster 
to $E(B-V)=1.22\pm0.22$ mag, corresponding to A$_V$=3.9$\pm$0.7 mag
(assuming R$_V$=3.2). 

\subsection{Observations and Data Reduction}

Infrared imaging 
observations were carried out on Jan 16, 2003, with the Isaac
Newton Group Red Imaging Device (INGRID) at the Cassegrain focus of
the 4.2m WHT. The instrument uses 1024x1024 Hawaii near-IR detector
array. The scale is 0.238 arcsec $\rm pixel^{-1}$ giving a total
field of view of 4.06x4.06 arcmin. We obtained two sequences of
shallow and deep $JHK_S$ images, to extend the dynamical range of
the photometry. The former series consists of 5 (6 for the $J$ band)
1 sec dithered images, alternated with off-cluster images for sky
subtraction. The deep series follows the same pattern but the
individual images consist of 6 averaged 10 sec frames (12 $\times$ 
5 sec for the $K_S$ band).

The stellar photometry of the combined and sky-subtracted images was 
carried out using ALLSTAR in DAOPHOT II (Stetson \cite{ste93}). 
We consider only stars with DAOPHOT errors less than 0.2 mag. 
The median averaged photometric errors are $0.02\pm0.01$ for the $J,H,K_s$ 
magnitudes 
brighter than 17 mag and $0.04\pm0.03$ for the fainter stars. There is 
also an additional observational uncertainty of $\sim$0.03 mag due to 
the sky background variations. The six brightest stars $(K_s < 12)$ 
are saturated in our photometry. We used instead the 2MASS data. 

The photometric calibration was performed by comparing our instrumental 
magnitudes with the 2MASS magnitudes of 20-25 stars (depending on the 
band). The transformation equations are:

\begin{eqnarray*}
  J&=&(0.99\pm0.02)\times j+(0.02\pm0.01)\times (j-k_s)-\\
&& (2.27\pm0.20)\hspace{10pt} (r.m.s.=0.06)\\
  H&=&(0.98\pm0.02)\times h-(0.01\pm0.01)\times (h-k_s)-\\
&&(1.92\pm0.16)\hspace{10pt} (r.m.s.=0.04)\\
  K_s&=&(1.01\pm0.01)\times k_s+(0.02\pm0.01)\times (j-k_s)-\\
&&(2.56\pm0.15)\hspace{10pt} (r.m.s.=0.03)\\
\end{eqnarray*}
where $j,h,k_s$ are the instrumental magnitudes, and $J,H,K_s$ are 
the magnitudes in the 2MASS system. The standard error values for 
the coefficients are also given. 
 
Our final photometric list contains 733 stars with location on the 
final images in pixels coordinates (x,y) and $J$, $H$ and $K_s$ 
magnitudes and their photometric uncertainties. The complete data 
set is available in the electronic edition of the Journal. The 
artificial star technique (Stetson \& Harris \cite{ste88}; Stetson 
\cite{ste91a}, \cite{ste91b}) was used to determine the $100\%$
completeness limits of the data: $(J, H, K_s)_{lim} =
 (16.5, 17.5, 17)$ mag.

\subsection{Star Counts: Location and Boundaries of the Cluster}

The object has a complex morphology, as becomes evident from the 
true-color image of the region (Fig.~\ref{fig01}). Carpenter et al. 
(\cite{car93}) reported a cluster that contained several bright stars 
associated with the H{\sc ii} region Sh\,2-228. Our data revealed a 
rich cluster of fainter stars, which are not visible on their 
images, and are only partially visible on the 2MASS (Paper {\sc I}). 
It is no surprise since the formation of 
high-mass stars is often related to the formation of a cluster of 
less massive stars (i.e. Tapia et al. \cite{tap91}, \cite{tap96}; 
Lada \& Lada \cite{lad91}; Persi et al. \cite{per94}, \cite{per00}).

\begin{figure}
  \caption{Near infrared 3-color composite image of \object{CC\,01}. 
  The $J$, $H,$ and $K_s$ bands are mapped onto blue, green and red, 
  respectively. The image is 4.06 arcmin on the side. North is up, 
  and East is to the left.}
  \label{fig01}
\end{figure}

To determine the location and the boundaries of the cluster we 
performed direct star counts, assuming spherical symmetry. We
varied the center searching for the steepest radial profile.
The new cluster coordinates are: $\alpha_{2000}$=05:13:28 
$\delta_{2000}$=+37:26:53, with uncertainty of about 10 arcsec, 
which are shifted by $\sim$0.4 arcmin to the South-East comparing 
with the ones determined by Carpenter et al. (\cite{car93}) and 
Paper {\sc I}. 

\begin{figure}[htbp]
\resizebox{\hsize}{!}{\includegraphics{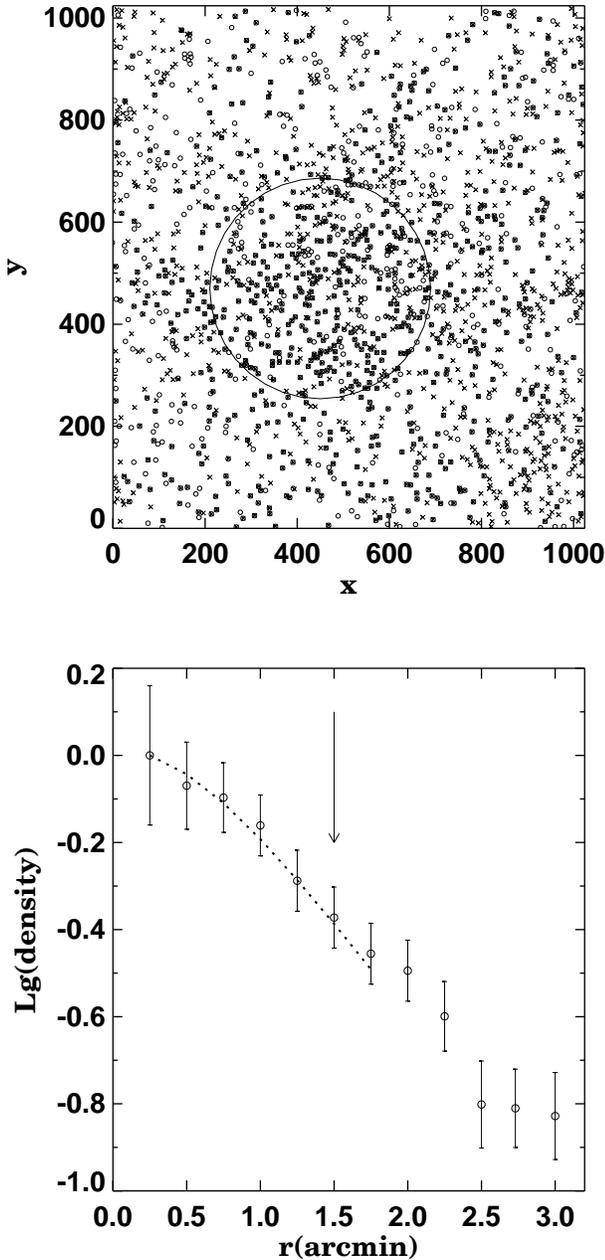}}
\caption{Cluster morphology: Map of stars detected in $K_s$ 
(circles) and $H$ (crosses). 
The large open circle marks the boundaries of the cluster assuming 
spherical symmetry.  Radial profile of the projected $K_s$ star number density 
(in number of stars per square arcmin) versus the radius. Bars are 
$1\sigma$ uncertainties from the Poisson statistics. The best fitting 
King profile (dashed line) is shown for comparison. The arrow indicates 
a radius of 1.5 arcmin (see Sec.~4.2).
}
\label{fig02}
\end{figure}

The cluster boundary was defined, perhaps conservatively, as the 
point of the profile where the excess density becomes twice the 
standard deviation of the surface density of the surrounding field. 
This yields a radius of 1.5 arcmin, indicated by a circle in the 
upper panel of Fig.~\ref{fig02}. Note that the cluster might be
larger as the radial profile flattens only at $\sim$2.3 arcmin but
the excess is below the 2$\sigma$ significance this far. 

Careful inspection of our deep images shows some cluster 
asymmetry -- there is an excess of faint stars to the East -- 
South-East (at $\alpha_{2000}$=05:13:34 $\delta_{2000}$=+37:26:46; 
Fig.~\ref{fig01}). The stellar surface density there is $122\pm10$ 
stars per sq. arcmin, significantly higher than the background of 
$37\pm5$ stars per sq. arcmin, measured in the member-free 
North-East region. This reveals an asymmetric and elongated 
cluster. Even on the infrared images however (Fig.~\ref{fig01})
is evident that there is a big absorption gradient, along the
NW-SE direction and the true shape of the cluster cannot be
completely defined, until the cluster emerges from 
its obscure cloud.

Fig.~\ref{fig02} shows the positions of all stars measured in the 
$H$ and $K_s$ bands (top) and projected $K_s$ band star number 
density in number of stars per sq. arcmin (bottom). The best 
fitting King profile ($r_c=1.5$ and $r_t=7$ arcmin, c=0.67) is 
shown just for comparison, because there is no physical reason to 
believe, that such an young cluster should follow a King profile.

\subsection{Color-Color and Color-Magnitude Diagrams}

\begin{figure}[htbp]
\resizebox{\hsize}{!}{\includegraphics{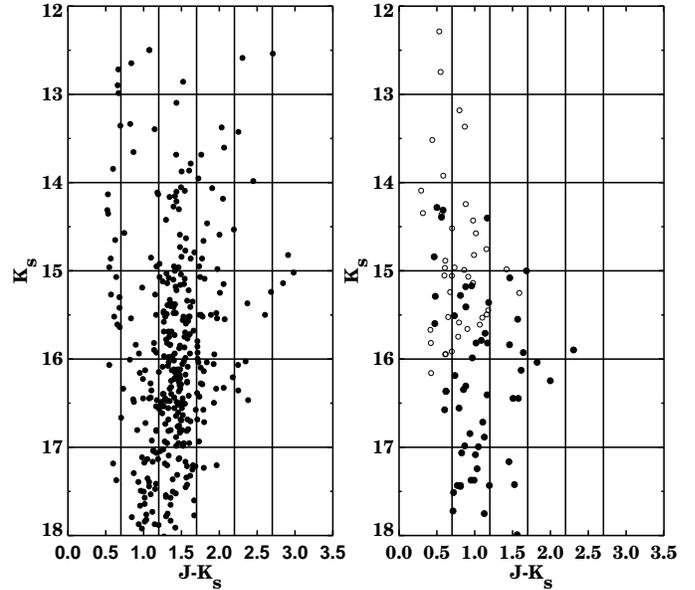}}
\caption{The $(J-K_s,K_s)$ color-magnitude diagram for CC\,01 + field stars 
within $90\arcsec$ (left panel). The field CMD (right-hand-side panel)
is based on all stars in the normalized area with $r\ge 2 \arcmin$ (see text). 
Second field from 2MASS is shown for comaprison with open circles.
} 
\label{Fig03}
\end{figure}

To decontaminate the cluster's CMD we used above determined cluster 
limiting radius $r=1.5\arcmin$ encircling the most probable cluster members. 
This area contains cluster's stars + field stars. 
The field stars alone are selected in an annulus around the
cluster with inner radius $r = 2.0 \arcmin$ and normalized to the same
area covered by the CC\,01. 
To check the location of field stars on the CMD we retrieved 
the photometry from 2MASS for a second comparison field situated 10 
arcmin from CC\,01 (open circles on (Fig.~\ref{Fig03})).
The $(J-K_s,K_s)$ CMDs of ``cluster + field" 
and ``field" are  gridded and the stars in each box in the two 
diagrams are counted (Fig.~\ref{Fig03}). 
Then an equivalent number of stars is removed from the single boxes of 
the ``cluster + field" CMD on the basis of the number of field stars found 
in the ``field" CMD alone.

%

\begin{figure}[htbp]
\resizebox{\hsize}{!}{\includegraphics{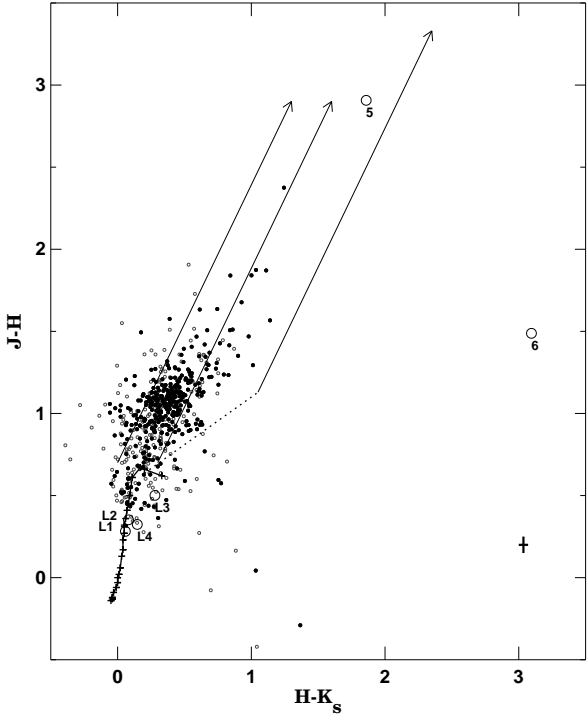}}
\caption{The $H-K_s$ vs. $J-H$ color-color diagram of all 
stars in our photometry list. The cluster members are solid dots, 
and the field stars are open circles. The labeled large open circles 
indicate the six brightest stars with 2MASS photometry.
The cross indicates typical photometric errors.
The Main Sequence line (Schmidt-Kaler \cite{sch82}) is shown with a solid 
line. Three reddening vectors (Bessell et al. \cite{bes98})  
are shown, which correspond 
to the visual extinction of 20 magnitudes. The two on the left encompass 
the main sequence stars. The unreddened location of T Tauri stars 
(Meyer et al. \cite{mey97}) 
is indicated as a dashed line, and the rightmost reddening vector 
brackets the locus of the reddened T Tau stars. 
}
\label{fig04}
\end{figure}

\begin{figure}[htbp]
\resizebox{\hsize}{!}{\includegraphics{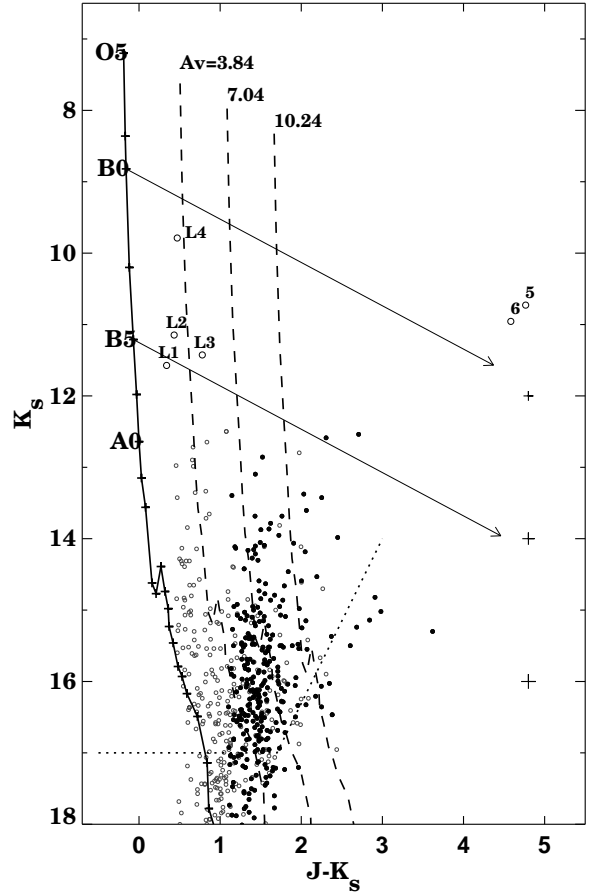}}
\caption{The $J-K_s$ vs. $K_s$ color-magnitude diagrams of 
all stars from our photometry list. Symbols are the same as in 
Fig.~\ref{fig04}. The Main Sequence
line (Schmidt-Kaler \cite{sch82}) is shown with solid line, the 
dashed lines stand for Main Sequence reddened with E($B-V$)=1.2, 
E($B-V$)=2.2 and E($B-V$)=3.2 mag. The reddening vectors correspond 
to A$_V$=20 mag. Dotted lines represent the completeness limit of 
the photometry. The $(m-M) = 11.99$  is adopted
from (Hunter \& Massey (\cite{hun90}). 
}
\label{fig05}
\end{figure}

The $J-H$ vs. $H-K_s$ color-color diagram is
 show in 
Fig.~\ref{fig04}. The cluster members are solid dots, and the field 
stars are open circles. As discussed above, we assume that the 
cluster is confined within 1.5 arcmin from the center and is
statistically decontaminated for the field stars.  
Three reddening vectors are shown. The two on the left encompass 
the main sequence stars, whose unreddened sequence
(Schmidt-Kaler \cite{sch82}) is represented as a solid line with crosses. 
The unreddened location of T Tauri stars (Meyer et al. \cite{mey97}) 
is indicated as a dashed line, and the rightmost reddening vector
brackets the locus of the reddened T Tau stars. Throughout the paper 
we used the absorption ratios from Bessell et al. (\cite{bes98}, Table A6).
The six brightest stars with 2MASS photometry are shown as large open 
circles and are labeled from 1 to 6 (the label L stand for the 
stars with optical photometry of Lahulla (\cite{lah85})).

Clearly, Fig.~\ref{fig04} indicates that most of the stars are 
reddened main-sequence stars. Note that the color spread of cluster 
members is 0.1-0.2 mag -- comparable with the $3\sigma$ of the 
typical photometric errors of 0.05 mag, -- suggesting no differential 
extinction within the object at the level of $\Delta$(A$_V$)$\sim$1 
mag. 

We will use the $K_s$ vs. $J-K_s$ color-magnitude diagram shown in
Fig.~\ref{fig05} to estimate the extinction toward the cluster 
members. The basic symbols are the same as in Fig.~\ref{fig04}. To 
plot the theoretical Main Sequence line (Schmidt-Kaler \cite{sch82}) 
we used as a first approximation the distance of $(m-M)_0=11.99$ mag
(D=2.5 Kpc) determined by Hunter \& Massey (\cite{hun90}), and based 
on the spectroscopy and optical photometry of the stars L3 and L4. 
The dashed lines are reddened Main Sequences for E($B-V$)=1.2, 
E($B-V$)=2.2, and E($B-V$)=3.2 mag. The reddening vectors for B0V 
and B5V stars correspond to A$_V$=20 mag and R$_V$=3.2. We  
assumed that the brightest 
stars L1, L2, L3, and L4 are affected by a relatively small visual 
extinction - 2.5-5 mag, - compared with the fainter stars. A 
comparison with the location of ``field'' stars suggests 
that they are subject only to foreground extinction. In other words, 
the stellar winds from the most massive stars have already cleared 
out their own dust envelopes. The spectral type of L3 is B: and of L4 
is B0V (Hunter \& Massey \cite{hun90}). They are probably among the 
major contributors to the excitation of Sh\,2-228. Note that in our
following considerations we adopted for L3 the spectral types derived 
by Crampton et al. (\cite{cra78}) who give B2IV, since the 
classification of Hunter \& Massey (\cite{hun90}) -- B: -- is 
somewhat uncertain. 

Stars 5 ($\alpha_{2000}$=05:13:26.08 $\delta_{2000}$=+37:27:08.27) 
and 6 ($\alpha_{2000}$=05:13:25.7 $\delta_{2000}$=+37:27:10.7), 
located at the center of Sh\,2-228 are extremely 
reddened. They have 
$K_s$=10.73, $J-K_s$=4.77, $H-K_s$=1.86 and 
$K_s$=10.96, $J-K_s$=4.58, $H-K_s$=3.10 mag, respectively. 
If they are main sequence stars with 
E($J-K_s$)=4.93 and E($J-K_s$)=4.74 mag, 
they have visual extinctions of A$_V$=27.2 and A$_V$=26.2 mag. 
Star 5 could be also T Tauri star and in this case A$_V$=17.6 mag. 
And finally, this star could be an Herbig AB star.
Most of the cluster members occupy the locus between E($B-V$)=1.2 
and E($B-V$)=3.2 mag. Approximately 20$\%$ of them show indications 
of IR excess. We adopt as a mean reddening of the cluster
E($B-V$)=2.4.

\subsection{Photometric Distance}

An approximate distance estimate, was carried out, based on the 
optical spectral classification of individual stars, following the 
example of Hunter at al. (\cite{hun90}). However, we used the
infrared luminosity of the stars, minimizing the reddening effects.
For the stars with known spectral types (L3 and L4), we
compared our observed $J-K_s$ color 
with the intrinsic one (Schmidt-Kaler \cite{sch82}) to determine 
E($J-K_s$), A$_V$ and A$_{K_s}$. Next, we used the apparent 
$K_s$ and the absolute $M_{K_s}$ (Schmidt-Kaler \cite{sch82}) 
magnitudes for the given spectral class to derive the distance 
moduli: $(m-M)_0$=12.61 mag for L3 and $(m-M)_0$=12.51 mag for L4. 
Our final distance modulus is $(m-M)_0$=12.56$\pm$0.08 mag
(D=3.5$\pm$0.1 Kpc). The error includes the standard deviation 
of the mean value, the uncertainties of photometry, and the 
photometric calibration. A possible unaccounted sources of 
systematic errors are the unreddened colors but we have no way 
of estimating them. Our result falls into the limits (2.4-5 Kpc)
given by Hunter at al. (\cite{hun90}).

\subsection{Cluster Age}

\begin{figure}
\resizebox{\hsize}{!}{\includegraphics{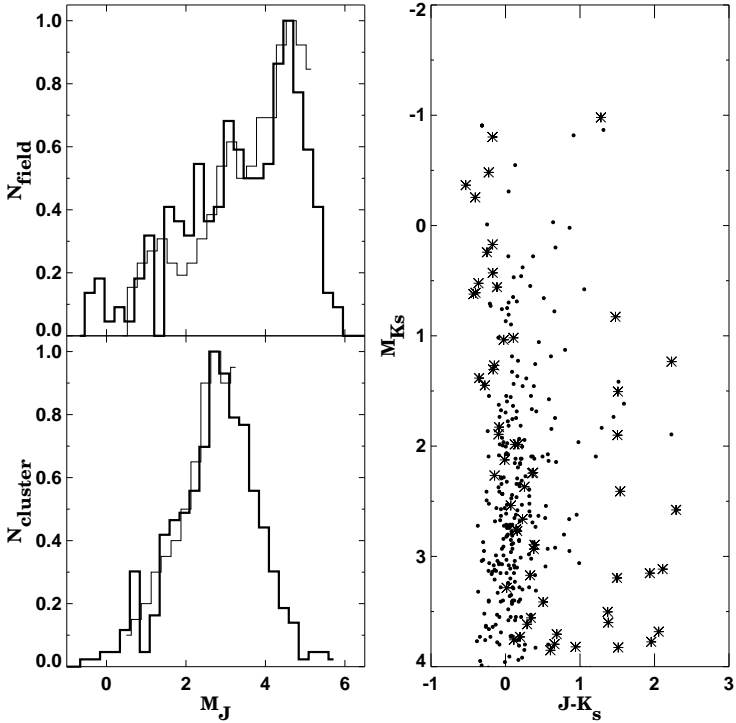}}
\vspace{4mm}
\caption{The observed luminosity function of the cluster (bottom 
left) and field stars (top left). The thin lines are the theoretical 
models by Strom et. al (\cite{sto93}) for 3 and 7 Myr, and the thick 
lines are the observed luminosity functions. 
A comparison of CC01 (solid dots) with S233IR (asterisk) from Porras 
et al. (\cite{por00}) is shown on the right panel. For CC\,01 we 
adopted E($B-V$)=2.4, and $(m-M)_0$=12.56 mag, and for S233IR -- the 
cluster distance of D=1.8 Kpc, and the individual reddening estimates 
from Porras et al. (\cite{por00}).
}
\label{fig06}
\end{figure}

We estimated approximately the age of the cluster and field stars 
by comparing the observed and theoretical luminosity functions (LF), 
closely following the method used by Porras et al. (\cite{por00}).
Strom et al. (1993) presented an evolutionary sequence of six 
$J$ band LFs for 0.3, 0.7, 1, 3 and 7 Myr. We used E($B-V$)=2.4 and 
E($B-V$)=1.2 mag for the cluster and the filed stars, respectively, 
and $(m-M)_0$=12.56 mag. We applied the Kolmogorov-Smirnov
test to the observed and 
theoretical LFs, looking for the closest resemblance. With $75\%$ 
and $67\%$ the Kolmogorov-Smirnov favors 3 Myr for the cluster stars
and 7 Myr for field population. Our best eye-ball match determined
a similar age of 1-2 Myr for the cluster population, and 5-6 Myr 
for the fields. 
Model predicted and observed $J$ band LF are presented in 
Fig.~\ref{fig06}. 

The color-magnitude diagram of CC\,01 is compared with the main 
sequence of S233IR with estimated age of $\sim$3 Myr (Porras et al. 
\cite{por00}) in the right panel of Fig.~\ref{fig06} . The main sequence stars in CC\,01
are shown with solid dots, and the S233 stars -- 
with asterisk. We adopted for CC\,01: E($B-V$)=2.4 
and $(m-M)_0$=12.56 mag. The parameters of S233IR are taken from 
Porras et al. (\cite{por00}; Table 1). As can be seen we have a good
agreement between our photometry and that 3 Myr young cluster.

\subsection{Initial Mass Function and Total Cluster Mass}

\begin{figure}
\resizebox{\hsize}{!}{\includegraphics{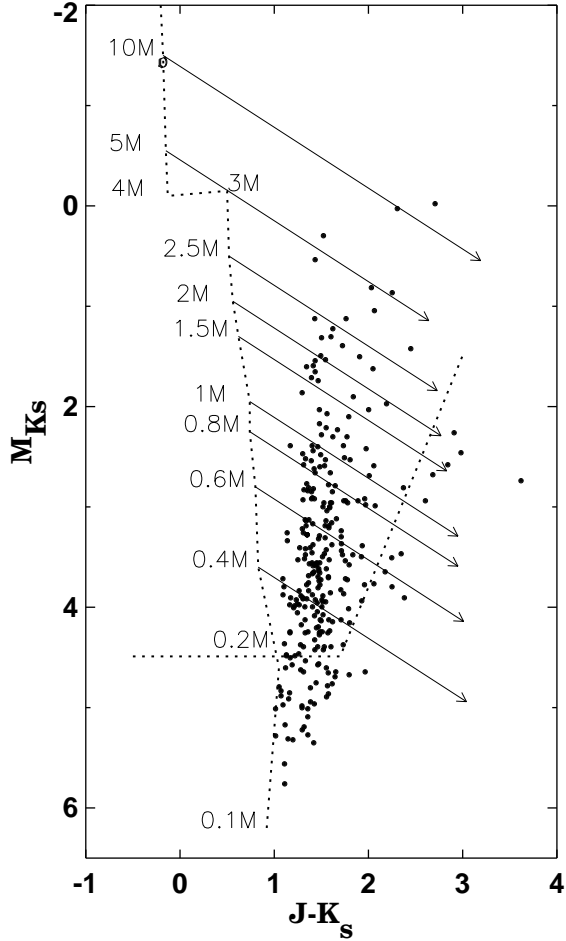}}
\caption{The $(J-K_s)_0$ vs. $M_{K_s}$ color-magnitude diagram
of CC\,01 members. The completeness limit is
shown as dotted line. The 3 Myr isochrone with stellar masses is
from pre-main sequence evolutionary tracks by D'Antona \& 
Mazzitelli (\cite{ant94}). Reddening vectors are from Bessell 
et al. (\cite{bes98}).
}
\label{fig07}
\end{figure}

\begin{figure}
\resizebox{\hsize}{!}{\includegraphics{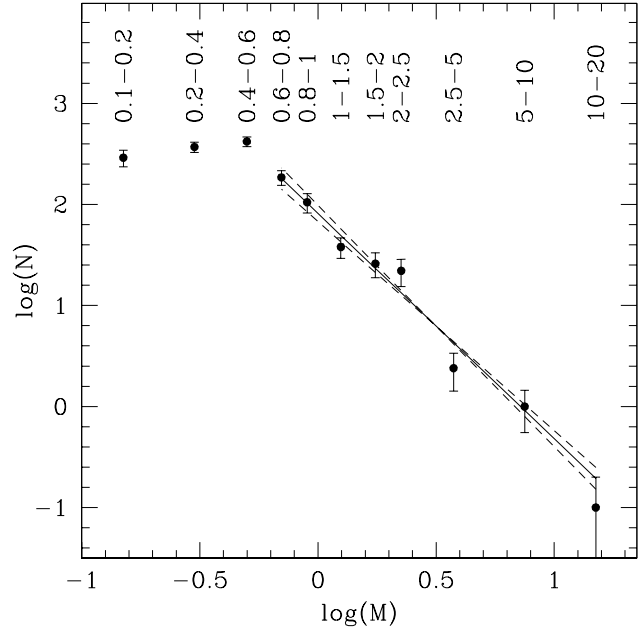}}
\caption{The initial mass function of the cluster. 
Error bars are calculated from the Poisson statistics of the 
number of stars in each bin. The $1\sigma$ uncertainty of the 
slope is shown with dashed lines. The limits of the bins in 
solar masses are indicated above. 
}
\label{fig08}
\end{figure}

Once the distance, the extinction, and the age of the cluster are
known, it is easy to determine approximately the slope of the 
initial mass function (hereafter IMF). 
We adopted the 3 Myr isochrone, and counted 
cluster stars between reddening lines, originating from positions 
on the isochrone for different initial masses (Fig.~\ref{fig07}). 
The region between 2.5 and 5 $M_\odot$ was treated as a single bin 
to avoid ambiguity. The derived slope over the range 0.6-20 $M_\odot$ 
is $-$2.23$\pm$0.16, close to the Salpeter (\cite{sal55}) slope of 
$-$2.35 (Fig.~\ref{fig08}). We used least square fit, weighted by 
the statistically uncertainly of the number of stars in each bin. The 
lower mass limit corresponds to the completeness limit of our 
photometry. However, the excellent agreement between the number of 
stars in the 0.4-0.6 $M_\odot$ bin and the IMF 
suggest that our completeness limit might be conservative.  

This estimate of the IMF slope is an 
approximation because of both the observational errors, and the 
systematic uncertainties related to the isochrones, or 
model-generated luminosity functions. Only spectroscopic 
observations allow to obtain accurate stellar masses. 

Finally, we estimated the total cluster mass by integrating the 
IMF over the entire range from 0.1 to 20$M_\odot$, 
to obtain $M_{total}\lesssim 1800\pm 200 M_\odot$. The uncertainty -- 
which takes into account the error in the slope, and the distance 
to the cluster -- is dominated by the extrapolation toward the low 
mass stars.

We removed the fore- and background contribution only statictically
and we extrapolated the IMF 
toward low mass stars linearly (in log-log space) while it has been 
shown that the IMF flattens or turns over at 
$\sim$0.8$M_\odot$ (i.e. Brice\~{n}o et al. \cite{bri02}). These 
effects are irrelevant for the goal of this work because we are 
searching for massive clusters, and therefore, an upper mass limit 
is sufficient. In fact, our extrapolation produces an overestimate 
nearly by a factor of two because about half of the total mass 
derived above falls into the 0.1-0.2 $M_\odot$ bin. The lower mass
limit is also somewhat debatable, but the flattening of the  
mass function renders the uncertainty negligible.

\section{Discussion and Summary}

We report three new obscured Milky Way clusters discovered using our 
algorithm that searches for peaks in the stellar surface density
histogram of the 2MASS point source catalog. This search completes 
the 5x5 arcmin bin survey. One more cluster was discovered 
serendipitously during a visual inspection of the candidates. In 
most of the cases the 2MASS photometry is insufficiently deep to 
assess the nature of the clusters. Follow-up near infrared imaging of 
some objects is planned.

We also present the first deep $J$, $H$, and $K_s$ imaging of the 
cluster CC\,01. The goal of this project -- to search for
massive obscured clusters, analogs of the Arches, -- makes it vital
to derive the total mass or at least an upper limit to it. To 
achieve this we estimated the cluster age to $\sim$1-3 Myr, the 
distance modulus (m-M)$_0$=12.56$\pm$0.08 mag (D=3.5 Kpc), and the
extinction A$_V$$\sim$2.4 mag. Finally, we derived the slope of the 
IMF for clusters stars: $\Gamma$=$-$2.23$\pm$0.16, 
which is close to the canonical Salpeter slope of $-$2.35. The 
integration over the IMF yielded an upper limit to 
the total cluster mass of $M_{total}^{max}$=1800$\pm$200$M_\odot$. 

CC\,01 appears to be a regular ``run of the mill'' star formation 
region, including the asymmetrical shape. For example, it is very 
similar to S233IR cluster (Porras et al \cite{por00}) and the 
clusters associated with H{\sc ii} regions Sh\,217 and Sh\,219 
(Deharveng et al. \cite{deh03a}). All they contain young stars with 
an upper mass limit of $\sim$10 $M_\odot$, and are located at the 
periphery of the H{\sc ii} regions. The presence of associated mid 
infrared, radio, and/or X-ray sources is also typical for this 
class of objects. 

Deharveng et al. (\cite{deh03b}) suggested that such companion 
clusters can be formed in the framework of the ``collect and 
collapse'' model (Elmegreen \& Lada \cite{elm77}; Whitworth et al. 
\cite{whi94}; Elmegreen \cite{elm98}). This model predicts that when 
an H{\sc ii} region expands, the dense neutral material accumulates 
between the ionization front and the shock front which precedes it 
on the neutral side. The decelerating shocked layer can become 
unstable and collapse, as it is probably the case of CC\,01. Thus, 
CC\,01 can well be another example of second-generation cluster, 
formed recently from the surrounding H{\sc ii} region. 

\begin{acknowledgements}
This publication makes use of data products from the Two Micron
All Sky Survey, which is a joint project of the University of
Massachusetts and the Infrared Processing and Analysis
Center/California Institute of Technology, funded by the National
Aeronautics and Space Administration and the National Science
Foundation. This research has made use of the SIMBAD database,
operated at CDS, Strasbourg, France.
J.B and V.I thanks to F. Cameron and M. Catelan for useful comments. 
The authors gratefully acknowledge 
the comments by an anonymous referee. 

J.B. is supported by FONDAP 
Center for Astrophysics grant number 15010003. 

\end{acknowledgements}

\end{document}